\documentclass{article}
\usepackage[preprint]{spconfa4}

\usepackage{amssymb,amsmath,amsfonts}
\usepackage{bm}
\usepackage{booktabs}
\usepackage{cite}
\usepackage{comment}
\usepackage[shortcuts]{extdash}
\usepackage{graphicx}
\usepackage{microtype}
\usepackage{multirow}
\usepackage{url}
\usepackage{lipsum}

\usepackage{xargs}                      %
\usepackage[pdftex,dvipsnames]{xcolor}  %
\usepackage[colorinlistoftodos,prependcaption]{todonotes}
\newcommandx{\unsure}[2][1=]{\todo[linecolor=red,backgroundcolor=red!25,bordercolor=red,#1]{#2}}
\newcommandx{\change}[2][1=]{\todo[linecolor=blue,backgroundcolor=blue!25,bordercolor=blue,#1]{#2}}
\newcommandx{\info}[2][1=]{\todo[linecolor=OliveGreen,backgroundcolor=OliveGreen!25,bordercolor=OliveGreen,#1]{#2}}
\newcommandx{\improvement}[2][1=]{\todo[linecolor=Plum,backgroundcolor=Plum!25,bordercolor=Plum,#1]{#2}}
\newcommandx{\thiswillnotshow}[2][1=]{\todo[disable,#1]{#2}}

\usepackage{colortbl}
\usepackage{pgfplotstable}
\usepackage{xintexpr}

\usepackage{stackengine}
\usepackage{scalerel}
\newcommand\openbigstar[1][0.7]{%
  \scalerel*{%
    \stackinset{c}{0pt}{c}{}{\scalebox{#1}{\color{white}{$\bigstar$}}}{%
      $\bigstar$}%
  }{\bigstar}
}
\newcommand\openbigstarwhite[1][0.7]{%
  \scalerel*{%
    \stackinset{c}{0pt}{c}{}{\scalebox{#1}{\color{black}{$\bigstar$}}}{%
      $\bigstar$}%
  }{\bigstar}
}

\pgfplotsset{compat=1.16}

\usepackage{amsmath,amsfonts,amssymb,bm}

\newcommand{\mG}{\mathbf{G}}

\newcommand{\mW}{\mathbf{W}}
\newcommand{\mX}{\mathbf{X}}

\newcommand{\vu}{\mathbf{u}}

\newcommand{\vw}{\mathbf{w}}
\newcommand{\vx}{\mathbf{x}}

\copyrightnotice{978-1-6654-6867-1/22/\$31.00~\copyright2022 IEEE}
                   
\title{
    DNN-Free Low-Latency Adaptive Speech Enhancement Based on \\
    Frame-Online Beamforming Powered by Block-Online FastMNMF
}

\name{
    Aditya Arie Nugraha$^{1}$ \kern0.2em
    Kouhei Sekiguchi$^{1}$ \kern0.2em
    Mathieu Fontaine$^{3,1}$ \kern0.2em
    Yoshiaki Bando$^{4,1}$ \kern0.2em
    Kazuyoshi Yoshii$^{2,1}$
    \thanks{
        This work was supported by 
            JSPS KAKENHI Nos.~19H04137, 20K19833, and 20K21813.
    } %
}
\address{
$^{1}$Center for Advanced Intelligence Project (AIP), RIKEN, Japan \\
$^{2}$Graduate School of Informatics, Kyoto University, Japan \\
$^{3}$LTCI, Télécom Paris, Institut Polytechnique de Paris, France \\
$^{4}$National Institute of Advanced Industrial Science and Technology (AIST), Japan
} 

\begin{document}
\setlength{\abovedisplayskip}{5pt plus 1pt minus 1pt}
\setlength{\belowdisplayskip}{5pt plus 1pt minus 1pt}
\setlength{\abovedisplayshortskip}{5pt plus 1pt minus 1pt}
\setlength{\belowdisplayshortskip}{5pt plus 1pt minus 1pt}
\allowdisplaybreaks[4]

\setlength{\textfloatsep}{10pt plus 2pt minus 2pt}
\setlength{\floatsep}{8pt plus 2pt minus 2pt}
\setlength{\intextsep}{10pt plus 2pt minus 2pt}

\maketitle

\begin{abstract}
This paper describes a practical dual-process speech enhancement system 
 that adapts environment-sensitive frame-online beamforming (front-end)
 with help from environment-free block-online source separation (back-end).
To use minimum variance distortionless response (MVDR) beamforming, 
 one may train a deep neural network (DNN)
 that estimates time-frequency masks used for computing
 the covariance matrices of sources (speech and noise).
Backpropagation-based run-time adaptation of the DNN was proposed 
 for dealing with the mismatched training-test conditions.
Instead, one may try to directly estimate the source covariance matrices
 with a state-of-the-art blind source separation method called
 fast multichannel non-negative matrix factorization (FastMNMF).
In practice, however, neither the DNN nor the FastMNMF
 can be updated in a frame-online manner
 due to its computationally-expensive iterative nature.
Our DNN-free system leverages
 the posteriors of the latest source spectrograms given by block-online FastMNMF 
 to derive the current source covariance matrices for frame-online beamforming.
The evaluation shows
 that our frame-online system can quickly respond to 
 scene changes caused by interfering speaker movements
 and outperformed
 an existing block-online system with DNN-based beamforming
 by 5.0 points in terms of the word error rate.
\end{abstract}

\vspace{-.25\baselineskip}
\section{Introduction}
\label{sec:introduction}
\vspace{-.2\baselineskip}

\begin{figure}[t!]
    \centering\includegraphics[width=.98\linewidth]{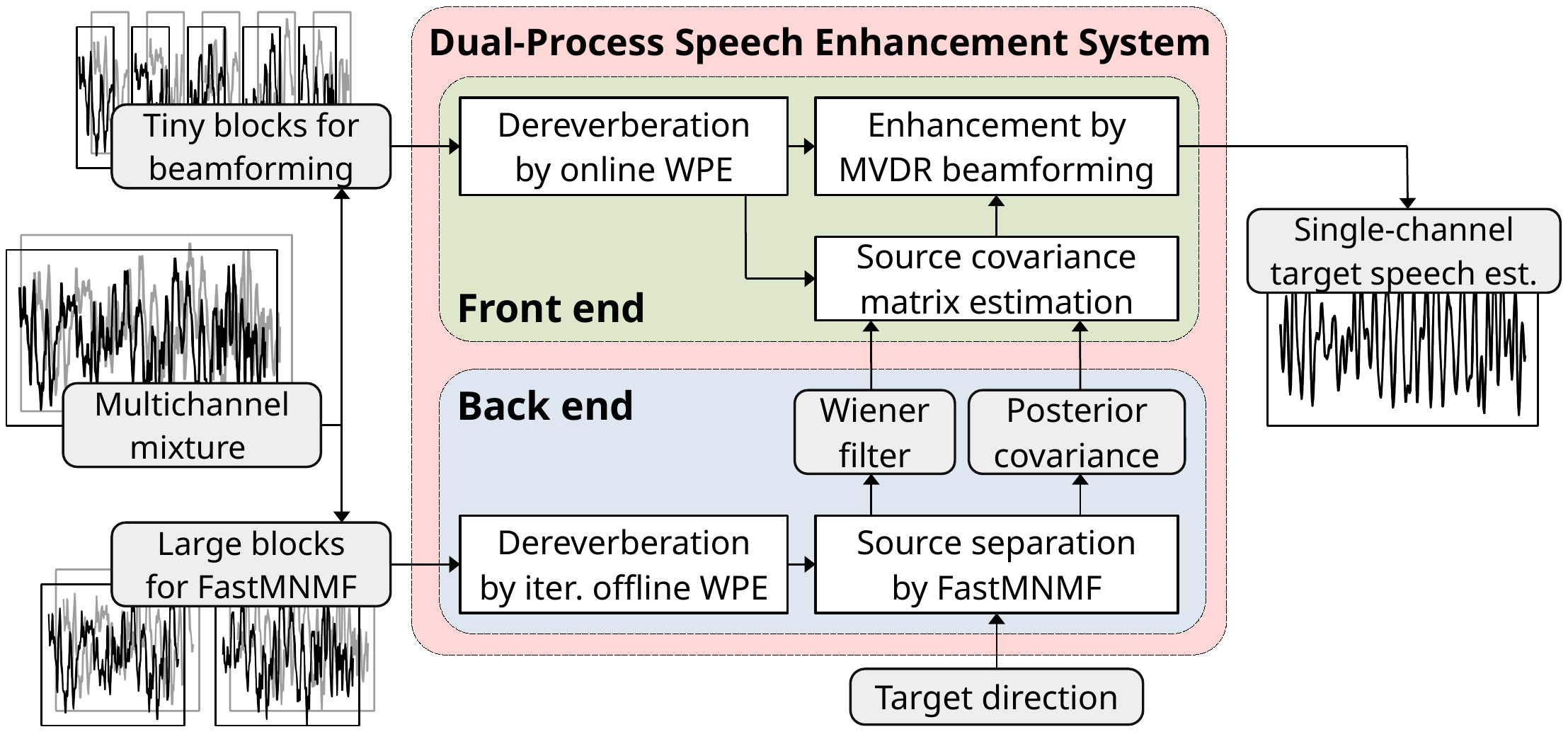}
    \vspace{-.75\baselineskip}
    \caption{The proposed low-latency speech enhancement system consisting 
    of a frame-online front end (beamforming) 
    informed by a block-online back end (FastMNMF).}
    \label{fig:proposed_system}
\end{figure}

In real environments,
speech enhancement methods must be adaptive to variations in sound scenes
caused by environmental changes or movements of the sound sources
\cite{alameda18multimodal,jurafsky21book,vincent18book}.
While it is important to successfully extract the speech of interest,
having a low computational cost can be critical for downstream tasks that demand low-latency outputs,
such as automatic speech recognition (ASR) for augmented reality applications
aiming at natural human-machine interaction.

Beamforming is a computationally-efficient multichannel source separation technique
that can extract a single-channel signal coming from a target direction
when an accurate steering vector or well-estimated source covariance matrices are given
\cite{vincent18book,kumatani12micarray,souden10optimal,heymann17neuralbf}.
Deep neural networks (DNNs) have been popular for estimating these source covariance matrices \cite{heymann17neuralbf,nugraha17thesis,sivasankaran20slogd,gu21complex,casebeer21nicebeam,sekiguchi22adaptive},
but these DNNs may have limited performance when the actual test environment is not covered by the training data.
In contrast, multichannel blind source separation (BSS) methods, such as
multichannel non-negative matrix factorization (MNMF) \cite{ozerov09mnmf}
and FastMNMF \cite{ito19fastmnmf,sekiguchi20fastmnmf},
are expected to perform well in any environment %
by optimizing the parameters of the assumed source probability distributions.
However, these methods require sufficient data and have a relatively high computational cost.

Semi-blind or blind source separation method has been combined with beamforming.
Beamformers have been derived in a block-online processing manner using 
the target source mask estimated as the posterior of
a complex Gaussian mixture model (cGMM) \cite{higuchi17cgmm} or
a complex angular central Gaussian mixture model (cACGMM) \cite{boeddecker18chime},
or the source covariance matrices obtained by MNMF \cite{shimada19unsupervised}.
These systems run the estimation of mask or covariance matrices and the beamforming sequentially on the same block of data.
Thus, the system latency is limited by the block size required by cGMM, cACGMM, or MNMF to provide reliable estimates.
FastMNMF, which has been shown to outperform MNMF,
has also been used as the back end of a dual-process adaptive online speech enhancement system \cite{sekiguchi22adaptive}
whose front end runs minimum variance distortionless response (MVDR) beamforming \cite{souden10optimal} with a DNN mask estimator \cite{heymann17neuralbf}.
The DNN mask estimator is periodically updated using the separated speech signals obtained by FastMNMF,
which carry out informed source separation given the target directions,
to adapt to the test environment.

This paper proposes a dual-process speech enhancement system
whose front end 
performs a responsive adaptive online MVDR beamforming %
by exploiting \textit{the parameters of the source posterior distributions},
i.e., the Wiener filters and the covariance matrices, estimated by the back-end FastMNMF,
as shown in Figure \ref{fig:proposed_system}.
Using those Wiener filters and source posterior covariance matrices,
we compute frame-wise second-order raw moments given the current observed mixture
and accumulate them using exponential moving averages (EMAs)
to obtain block-wise source covariance matrices for the beamformer.
Consequently, the estimation of the source covariance matrix relies on FastMNMF,
instead of a DNN-based mask estimator \cite{sekiguchi22adaptive}.
Directly using the average source covariance matrices estimated by FastMNMF in a way similar to \cite{shimada19unsupervised}
is also possible.
However,
when the back end processes a significantly larger data block than the front end,
the front end processes many blocks using the same beamformer
while waiting for the back end to provide new source covariance matrices.
Our proposed system is thus more preferable because it can promptly respond to the sound scene changes.

The evaluation used multiple sequences of mixtures,
in which the interfering speaker locations are different in separate mixtures.
Our system outperformed DNN-based beamforming \cite{sekiguchi22adaptive} in terms of word error rate (WER)
by 5.0 points using frame-online processing with a total latency of 22 ms.

\vspace{-.4\baselineskip}
\section{Proposed System}
\label{sec:proposed}
\vspace{-.3\baselineskip}

Let 
$\smash{\vx_{ft} \!\in\! \mathbb{C}^{M}}$ be
the short-time Fourier transform (STFT) coefficients at frequency $f \!\in\! [1, F]$ and time frame $t \!\in\! [1, T]$
of the observed multichannel mixture signal captured by $M$ microphones
and
$\smash{\vx_{nft} \!\in\! \mathbb{C}^{M}}$ be
the STFT coefficients of the so-called multichannel image of source $n \!\in\! [1, N]$,
where $F$ is the number of frequency bins, $T$ is the total number of time frames,
and $N$ is the number of sources.
The source images are assumed to sum to the observed mixture as
$\smash{\vx_{ft} \!=\! \sum_{n=1}^N \vx_{nft}}$.
Given $\mX \!\triangleq\! \left\{\vx_{ft} | \forall f,\! \forall t \right\}$,
BSS generally estimates $\forall n, \allowbreak \mX_n \!\triangleq\! \left\{\vx_{nft} | \forall f,\! \forall t \right\}$.
In this paper, our dual-process adaptive online speech enhancement system
obtains the single-channel signal estimate $\left\{s_{n^\prime ft} | \forall f,\! \forall t \right\}$ of target source $n^\prime$ for ASR purpose.

The dual-process speech enhancement system executes
a back end (Sect.~\ref{sec:proposed_backend}) and 
a front end (Sect.~\ref{sec:proposed_frontend}) in parallel
in a block-online processing manner,
where each block is a subset of $\mX$ consisting of a sequence of multiple time frames, as in \cite{sekiguchi22adaptive}.
At a time,
the back end processes $\smash{\mX^\text{BSS}_i} \!\subset\! \mX$
composed of $\smash{T^\text{BSS}}$ frames with $i$ is the block index,
while the front end processes $\smash{\mX^\text{BF}_j} \!\subset\! \mX$
composed of $\smash{T^\text{BF}}$ frames with $j$ is the block index.
$\smash{T^\text{BSS}}$ can be large enough to provide reliable statistics required for good BSS performance,
while $\smash{T^\text{BF}}$ can be small when low-latency outputs are expected.

The proposed system is similar to the system in \cite{sekiguchi22adaptive}.
Both systems basically use the same back-end FastMNMF.
However, the front ends use different ways
to obtain the source covariance matrices required to derive MVDR beamformer.

\vspace{-1.2\baselineskip}
\subsection{Back End}
\label{sec:proposed_backend}
\vspace{-.4\baselineskip}

As in \cite{sekiguchi22adaptive},
the back end operates given a block of data $\smash{\mX^\text{BSS}_i}$ and one or more target directions.
Offline iterative dereverberation \cite{drude18narawpe} is first performed on $\smash{\mX^\text{BSS}_i}$
to obtain a set of less reverberant mixtures $\smash{\widehat{\mX}^\text{BSS}_i}$,
on which FastMNMF \cite{sekiguchi20fastmnmf} is then applied
after initializing the inverse of the so-called diagonalization matrix given the target directions.
Although FastMNMF typically aims for the source image estimates,
we are more interested in the estimated parameters of the posterior distribution $\widehat{\vx}_{nft} | \widehat{\vx}_{ft}$.

The local Gaussian model
assumes that each source image $\widehat{\vx}_{nft}$ follows an $M$-variate complex\-/valued circularly\-/symmetric Gaussian distribution, 
whose covariance matrix is decomposed into power spectral density (PSD) $\lambda_{nft}$ and spatial covariance matrix (SCM) $\mG_{nf}$, as
$\widehat{\vx}_{nft} \!\sim\! \mathcal{N}_\mathbb{C}\left(\mathbf{0}, \lambda_{nft} \mG_{nf} \right)$ \cite{duong10underdetermined}.
To deal with the difficult optimization of this vanilla model,
the state-of-the-art BSS method called FastMNMF \cite{sekiguchi20fastmnmf}
uses a nonnegative matrix factorization (NMF)-based spectral model
and a jointly-diagonalizable spatial model.
The PSD is given by
$\smash{\lambda_{nft}} \!\triangleq\! \smash{\sum_{c=1}^C u_{ncf} v_{nct}} \!\in\! \smash{\mathbb{R}_{+}}$,
where $\smash{u_{ncf}} \!\in\! \smash{\mathbb{R}_{+}}$ and $\smash{v_{nct}} \!\in\! \smash{\mathbb{R}_{+}}$
with $c \!\in\! [1, C]$ and $C$ is the number of NMF components.
The SCM is jointly-diagonalizable by
a time-invariant diagonalization matrix shared among all sources
$\mathbf{Q}_f \!\in\! \smash{\mathbb{C}^{M \times M}}$ as 
$\mG_{nf} \!\triangleq\! \mathbf{Q}_f^{-1} \mathrm{Diag}(\tilde{\mathbf{g}}_{n}) \mathbf{Q}_f^{-\mathsf{H}}$,
where $\mathrm{Diag}(\tilde{\mathbf{g}}_{n})$ is a diagonal matrix
whose diagonal vector is $\smash{\tilde{\mathbf{g}}_{n}} \smash{\triangleq} \smash{\left[ \tilde{g}_{1n}, \dotsc, \tilde{g}_{Mn} \right]^\mathsf{T}} \!\in\! \smash{\mathbb{R}_{+}^{M}}$. 
Thus, the probability distributions of the $n$-th less reverberant source image and less reverberant mixture can be expressed as
\begin{align}
\widehat{\mathbf{x}}_{nft} &\!\sim\!
\mathcal{N}_\mathbb{C}^{M} \left(
    \mathbf 0,
    \lambda_{nft} \mathbf{Q}_f^{-1} \mathrm{Diag}(\tilde{\mathbf{g}}_{n}) \mathbf{Q}_f^{-\mathsf{H}}
\right) , \\
\widehat{\mathbf{x}}_{ft} &\!\sim\!
\mathcal{N}_\mathbb{C}^{M} \left(
    \mathbf 0,
    \sum\nolimits_{n=1}^N \lambda_{nft} \mathbf{Q}_f^{-1} \mathrm{Diag}(\tilde{\mathbf{g}}_{n}) \mathbf{Q}_f^{-\mathsf{H}}
\right) .
\end{align}
Consequently, the posterior distribution is given by
\begin{align}
    \widehat{\mathbf{x}}_{nft} \mid \widehat{\mathbf{x}}_{ft} &\sim \mathcal{N}_\mathbb{C}^M \big( \mathbf{W}_{nft} \widehat{\mathbf{x}}_{ft} \, , \, \mathbf{\Sigma}_{nft} \big) , \\
    \mathbf{W}_{nft} &= \mathbf{Q}_f^{-1}
    \mathrm{Diag} \left(
        \frac{\lambda_{nft} \tilde{\mathbf{g}}_{n}}
        {\sum_{n^{\prime}=1}^{N} \lambda_{n^{\prime}ft}\tilde{\mathbf{g}}_{n^{\prime}}} \right)
    \mathbf{Q}_f , \\
    \mathbf{\Sigma}_{nft} &= \left( \mathbf{I} - \mathbf{W}_{nft} \right) \mathbf{Q}_f^{-1}
    \mathrm{Diag} \left(
        \lambda_{nft} \tilde{\mathbf{g}}_{n} \right)
    \mathbf{Q}_f,
\end{align}
where $\mathbf{I}$ is the identity matrix. 
After the FastMNMF parameter optimization for $\smash{\widehat{\mX}^\text{BSS}_i}$ is finished,
we compute the exponential moving averages (EMAs),
$\smash{\widetilde{\mW}_{nfi}}$ and $\smash{\widetilde{\bm{\Sigma}}_{nfi}}$,
with $\smash{\alpha^\text{BSS}}\!\!=\!1$ for $i\!=\!1$,
to represent the Wiener filter and the posterior covariance matrix, respectively, of source $n$ in block $i$ as follows:
\begin{align}
    \widetilde{\mW}_{nfi} &\!=\!
        \frac{\alpha^\text{BSS}}{T^\text{BSS}} \sum\nolimits_{t^\prime=1}^{T^\text{BSS}} \mathbf{W}_{nft^\prime} + (1 - \alpha^\text{BSS}) \widetilde{\mW}_{nf(i-1)}, \\
    \widetilde{\bm{\Sigma}}_{nfi} &\!=\!
        \frac{\alpha^\text{BSS} }{T^\text{BSS}} \sum\nolimits_{t^\prime=1}^{T^\text{BSS}} \mathbf{\Sigma}_{nft^\prime} + (1 - \alpha^\text{BSS}) \widetilde{\bm{\Sigma}}_{nf(i-1)} .
\end{align}

\vspace{-.75\baselineskip}
\subsection{Front End}
\label{sec:proposed_frontend}

Given a block of data $\smash{\mX^\text{BF}_j}$, the front end performs
an online dereverberation to obtain a less reverberant mixture $\smash{\widehat{\mathbf{x}}_{ft}}$
that is then used by a beamforming to compute the target signal estimate $s_{nft}$.
The beamforming uses $\smash{\widetilde{\mW}_{nfi^\prime}}$ and $\smash{\widetilde{\bm{\Sigma}}_{nfi^\prime}}$,
where $\smash{i^\prime}$ is the index of the latest block $\smash{\mX^\text{BSS}_{i^\prime}}$ processed by the back end.

\vspace{-.75\baselineskip}
\subsubsection{Online Dereverberation}
\label{sec:proposed_frontend_dereverb}

We remove late reverberation from the mixture $\smash{\mathbf{x}_{ft}}$
using an online variant of WPE \cite{yoshioka12wpe,drude18narawpe}:
$\smash{\widehat{\mathbf{x}}_{ft}} \!=\! \smash{\mathbf{x}_{ft}} \!-\! \smash{\mathbf{H}^\mathsf{H}_{ft} \widetilde{\mathbf{x}}_{f(t-\Delta)}} \!\in\! \smash{\mathbb{C}^{M}}$,
where $\smash{\widetilde{\mathbf{x}}_{f(t-\Delta)}} \!\in\! \smash{\mathbb{C}^{M\!K}}$ stacks
$\left\{ \mathbf{x}_{ft^\prime} \middle| t^\prime \!\in\! [t\!-\!\Delta\!-\!K\!+\!1, t\!-\!\Delta] \right\}$ and
$\smash{\mathbf{H}_{ft}} \!=\! \smash{\mathbf{R}_{ft}^{-1} \mathbf{P}_{ft}} \!\in\! \smash{\mathbb{C}^{M\!K \times M}}$ is the WPE filter with
$\smash{\mathbf{R}_{ft}} \!\in\! \smash{\mathbb{C}^{M\!K \times M\!K}}$,
$\smash{\mathbf{P}_{ft}} \!\in\! \smash{\mathbb{C}^{M\!K \times M}}$,
$\Delta$ is the delay, and
$K$ is the number of filter taps.
Although our front end works in a block-online processing manner,
we opt for an online variant that allows us to avoid frequent matrix inversion $\smash{\mathbf{R}_{ft}^{-1}}$ when $T^\text{BF}$ is small.
We first initialize $\smash{\mathbf{R}_{f0}^{-1}} \!\leftarrow\! \mathbf{I}$ and $\smash{\mathbf{H}_{f0}} \!\leftarrow\! \mathbf{0}$,
where $\mathbf{0}$ is the zero matrix with appropriate dimensions. 
The dereverberation is then performed after updating $\smash{\mathbf{R}_{ft}^{-1}}$ and $\smash{\mathbf{H}_{ft}}$ as follows:
\begin{align}
    \phi_{ft} &\!=\! 
    (M\Delta)^{-1}
        \sum\nolimits_{\vphantom{A^\prime} m=1}^{\vphantom{A} M}
        \sum\nolimits_{\vphantom{A^\prime} t^\prime = (t - \Delta + 1)}^{\vphantom{A} t}
        \left| \left\{ \mathbf{x}_{ft^\prime} \right\}_m \right|^{2}, \\
    \mathbf{K}_{ft} &\!=\! \frac{
        \alpha^\text{WPE} \mathbf{R}_{f(t-1)}^{-1} \widetilde{\mathbf{x}}_{f(t-\Delta)}
    }{
        (1\!-\!\alpha^\text{WPE}) \phi_{ft} \!+\! \alpha^\text{WPE} \widetilde{\mathbf{x}}_{f(t-\Delta)}^\mathsf{H} \mathbf{R}_{f(t\!-\!1)}^{-1} \widetilde{\mathbf{x}}_{f(t-\Delta)}
    }, \! \\
    \mathbf{R}_{ft}^{-1} &\!=\!
    \left(1-\alpha^\text{WPE}\right)^{-1}\left(\mathbf{I}-\mathbf{K}_{ft}\widetilde{\mathbf{x}}_{f(t-\Delta)}^{\mathsf{H}}\right)\mathbf{R}_{f(t-1)}^{-1},
    \\
    \mathbf{H}_{ft} &\!=\! \mathbf{H}_{f(t-1)} + \mathbf{K}_{ft} \left(
        \mathbf{x}_{ft} - \mathbf{H}^\mathsf{H}_{f(t-1)} \widetilde{\mathbf{x}}_{f(t-\Delta)}
        \right)^\mathsf{H}, \label{eq:G_ft} \\
    \widehat{\mathbf{x}}_{ft} &\!=\! \mathbf{x}_{ft} - \mathbf{H}^\mathsf{H}_{ft} \widetilde{\mathbf{x}}_{f(t-\Delta)}. 
\end{align}

Our calculations for $\mathbf{K}_{ft}$ and $\mathbf{R}_{ft}^{-1}$ are slightly different from those presented in \cite{caroselli17adaptive,drude18narawpe,lemercier22endtoend} because we formulate EMAs:
$\mathbf{R}_{ft} \!=\!
\alpha^\text{WPE} \phi_{ft}^{-1} \widetilde{\mathbf{y}}_{f(t-\Delta)} \widetilde{\mathbf{y}}_{f(t-\Delta)}^\mathsf{H} \!+\!
(1\!-\!\alpha^\text{WPE}) \mathbf{R}_{f(t-1)}$ and
$\mathbf{P}_{ft} \!=\!
\alpha^\text{WPE} \phi_{ft}^{-1} \widetilde{\mathbf{y}}_{f(t-\Delta)} {\mathbf{y}}_{ft}^\mathsf{H} \!+\!
(1\!-\!\alpha^\text{WPE}) \mathbf{P}_{f(t-1)}$.
With these formulations, the EMA parameters in this paper, i.e., $\smash{\alpha^\text{BSS}}$, $\smash{\alpha^\text{WPE}}$, and $\smash{\alpha^\text{BF}}$,
provide the same interpretation about the weights of a new data and the accumulated data.
In terms of our $\smash{\alpha^\text{WPE}}$, the EMA parameter in \cite{lemercier22endtoend} is $\smash{(1\!-\!\alpha^\text{WPE})}$.

\vspace{-.75\baselineskip}
\subsubsection{Online Beamforming}
\label{sec:proposed_frontend_beamforming}

Assuming that $\widehat{\mathbf{x}}_{nft} \mid \widehat{\mathbf{x}}_{ft} \!\sim\! \smash{\mathcal{N}_\mathbb{C}^M \big( \widetilde{\mathbf{W}}_{nfi^\prime} \widehat{\mathbf{x}}_{ft} , \widetilde{\mathbf{\Sigma}}_{nfi^\prime} \big)}$,
we first compute the time-varying second-order raw moment as the covariance matrix $\smash{\bm{\Gamma}_{nft}}$ of source $n$
and the corresponding interference covariance matrix $\smash{\bm{\Upsilon}_{nft}}$.
EMAs $\smash{\widetilde{\bm{\Gamma}}_{nfj}}$, $\smash{\widetilde{\bm{\Upsilon}}_{nfj}}$
representing the source covariance matrices in block $j$
are calculated with $\smash{\alpha^\text{BF}}\!\!=\!1$ for $j\!=\!1$.
A beamformer $\smash{\vw^{\text{MV}}_{nfj}}$ \cite{souden10optimal} is then obtained
given a vector $\smash{\vu_{m^\prime}}$, whose ${m^\prime}$-th entry is $1$ and $0$ elsewhere
with ${m^\prime}$ is the reference microphone index, as follows:
\begin{align}
    \bm{\Gamma}_{nft} &= \widetilde{\mathbf{W}}_{nfi^\prime} \widehat{\vx}_{ft} \widehat{\vx}_{ft}^\mathsf{H} \widetilde{\mathbf{W}}_{nfi^\prime}^\mathsf{H} + \widetilde{\mathbf{\Sigma}}_{nfi^\prime}, \\
    \bm{\Upsilon}_{nft} &= \widehat{\mathbf{x}}_{ft} \widehat{\vx}_{ft}^\mathsf{H} - \bm{\Gamma}_{nft}, \\
    \widetilde{\bm{\Gamma}}_{nfj} &=
        \frac{\alpha^\text{BF}}{T^\text{BF}}
            \sum\nolimits_{t=1}^{T^\text{BF}} \bm{\Gamma}_{nft} +
        \left( 1 - \alpha^\text{BF} \right) \widetilde{\bm{\Gamma}}_{nf(j-1)}, \label{eq:gamma} \\
    \widetilde{\bm{\Upsilon}}_{nfj} &=
        \frac{\alpha^\text{BF}}{T^\text{BF}}
            \sum\nolimits_{t=1}^{T^\text{BF}} \bm{\Upsilon}_{nft} +
        \left( 1 - \alpha^\text{BF} \right) \widetilde{\bm{\Upsilon}}_{nf(j-1)}, \label{eq:upsilon} \\
    \vw^{\text{MV}}_{nfj} &=
        \Big( \text{tr} \big(
            \widetilde{\bm{\Upsilon}}_{nfj}^{-1} \widetilde{\bm{\Gamma}}_{nfj} \big) \Big)^{-1}
        \widetilde{\bm{\Upsilon}}_{nfj}^{-1} \widetilde{\bm{\Gamma}}_{nfj}
        \vu_{m^\prime}.
\end{align}
Finally, we use the beamformer for all time frames in block $j$, i.e., $\vw^{\text{MV}}_{nft} \!\leftarrow\! \vw^{\text{MV}}_{nfj}$,
to obtain a single-channel enhanced signal:
\begin{align}
    s_{nft} &=  \smash{ \big( \vw^{\text{MV}}_{nft} \big)}^\mathsf{H} \widehat{\vx}_{ft} .
\end{align}

\vspace{-.75\baselineskip}
\section{Evaluation}
\label{sec:evaluation}
\vspace{-.125\baselineskip}

This section presents the evaluation of our proposed method
on data recorded using a Microsoft HoloLens 2 (HL2).

\vspace{-.75\baselineskip}
\subsection{Experimental Settings}
\label{sec:exp_setting}

The evaluation was performed on the test subset of the dataset used in \cite{sekiguchi22adaptive}.
This subset contained eight simulated noisy mixture signals, each of which consists of multiple utterances, amounted to 18 min in total.
Each mixture signal was composed of two reverberant speech signals and one diffuse noise signal ($N\!=\!3$),
which were recorded separately in a room with an $\text{RT}_{60}$ of about 800 ms using the 5 microphones of an HL2 ($M\!=\!5$).
The dry speech signals were taken from the Librispeech dataset \cite{panayotov15librispeech},
and the noise signals were taken from the CHiME-3 dataset \cite{vincent17chime4}.
The noise source was located 3 m away from the HL2 in the direction of $135^{\circ}$, where $0^{\circ}$ was in front of the HL2,
behind multiple portable room dividers to build up reflections, which characterize a diffuse noise.
The target speaker and the interfering speaker were located 1.5 m away from the HL2.
The target speaker was in the direction of $0^{\circ}$, 
while the interfering speaker varied for each utterance between $\{45^{\circ}, 90^{\circ}, 180^{\circ}, 225^{\circ}, 270^{\circ}, 315^{\circ}\}$.
Each noisy mixture signal was fed in turn to a speech enhancement method,
so the method needs to handle the interfering speaker movements.

All audio signals were sampled at 16 kHz.
The STFT coefficients were extracted using a 1024-point Hann window ($F\!=\!513$) 
with 75\% overlap.
To factor out possible instability of the first few EMA computations due to improper initialization,
we concatenated the last 1024 frames ($\approx$ 16 s) of each noisy mixture to its beginning
so that the compared methods processed a few blocks before the performance measurement started.

The block size of the back end was set to $T^\text{BSS} \!=\! 256$ frames with 75\% overlap.
Thus, the back end provided new $\smash{\widetilde{\mW}_{nfi^\prime}}$ and $\smash{\widetilde{\bm{\Sigma}}_{nfi^\prime}}$ every 64 frames ($\approx$ 1 s).
The back-end iterative offline WPE was performed for 3 iterations using the tap length of 5 and the delay of 3.
The number of NMF components was $C\!=\!8$ and
the number of FastMNMF parameter updates was $50$, including $40$ warming-up iterations with the frequency-invariant source model.
The front-end online WPE was performed using the tap length of 5 and the delay of 3.
We loosely performed a grid search in preliminary experiments
by considering
$\alpha^\text{WPE}, \alpha^\text{BF}, \alpha^\text{BSS} \!\in\! \{ 0.500, \allowbreak 0.200, \allowbreak 0.100, \allowbreak 0.050, \allowbreak 0.020, \allowbreak 0.010, \allowbreak 0.005 \}$.
The experiments presented in this paper used $\alpha^\text{WPE} \!=\! 0.005$ and $\alpha^\text{BSS} \!=\! 0.100$.
The experimental results illustrate the proposed system's top performance on the test set
because the hyperparameter tuning for $\smash{\alpha^\text{WPE}}$, $\smash{\alpha^\text{BF}}$, and $\smash{\alpha^\text{BSS}}$ was performed on the same set.

The performance was evaluated in terms of the word error rate (WER) [\%] and the computation time [ms].
The ASR system was based on the transformer-based acoustic and language models of the SpeechBrain toolkit \cite{ravanelli21speechbrain}.
We additionally perform the standard statistical test
called the Matched Pair Sentence Segment Word Error (MAPSSWE) test \cite{gillick89stat}
to determine whether two WERs obtained by two different systems are different \cite{jurafsky21book}.
It is a two-tailed test whose null hypothesis is that
there is no performance difference between the two systems.
The computation time was measured on Intel Xeon E5-2698 v4 (2.20 GHz) with NVIDIA Tesla V100 SXM2 (16GB).

\vspace{-.75\baselineskip}
\subsection{Experimental Results and Discussion}
\label{sec:exp_result}

\begin{table}[t]
    \renewcommand{\arraystretch}{1.025}
    \setlength\tabcolsep{2.25pt}         %
    \setlength\aboverulesep{1pt}      %
    \setlength\belowrulesep{1pt}       %
	\vspace{-.5\baselineskip}
    \centering
    \caption{
        Average WERs [\%] and computation times [ms] of the \textit{baseline} front ends.
        The total latency [ms] is the sum of the block shift size and the average computation time for each block.
        Lower WER score and computation time are better.
    }
    \label{tab:baseline}
    \vspace{.1\baselineskip}
    \begin{tabular}{@{}l*{5}{c}@{}}
        \toprule
        & \multicolumn{2}{c}{Block} & & & \\
        \cmidrule(lr){2-3}
        \\
        \multirow{-3}{*}{Method} &
            \multirow{-2}{*}{\shortstack{\vphantom{bp} Size \\ \vphantom{bp} [ms]}} &
            \multirow{-2}{*}{\shortstack{\vphantom{bp} Shift \\ \vphantom{bp} [ms]}} &
            \multirow{-3}{*}{\shortstack{\vphantom{bp} Comp. \\ \vphantom{bp} Time \\ \vphantom{bp} [ms]}} &
            \multirow{-3}{*}{\shortstack{\vphantom{bp} Total \\ \vphantom{bp} Latency \\ \vphantom{bp} [ms]}} & 
            \multirow{-3}{*}{\shortstack{\vphantom{bp} WER \\ \vphantom{bp} [\%]}} \\
        \midrule
        Clean (ground truth) & --- & --- & --- & --- & $6.1$ \\
        Noisy (observation) & --- & --- & --- & --- & $92.1$ \\
        \midrule
        Online WPE & $16$ & $16$ & $3$ & $19$ & $87.8$ \\   %
        Online WPE + DS & $16$ & $16$ & $4$ & $20$ & $68.4$ \\   %
        Online WPE + MPDR & $16$ & $16$ & $6$ & $22$ & $47.1$ \\   %
        Online WPE + MPDR & $64$ & $64$ & $16$ & $80$ & $46.0$ \\   %
        Online WPE + MPDR & $256$ & $256$ & $54$ & $310$ & $47.2$ \\   %
        \midrule
        \multicolumn{6}{l}{\hspace{-.45em} WPE + MVDR with DNN-based mask estimation \cite{sekiguchi22adaptive}} \\
        \hspace{.45em} (before adaptation) & $3000$ & $500$ & $250$ & $750$ & $35.6$ \\
        \hspace{.45em} (after adaptation) & $3000$ & $500$ & $250$ & $750$ & $20.4$ \\
        \bottomrule
    \end{tabular}
\end{table}

Table \ref{tab:baseline} shows the baseline performances.
The WERs for `clean', `observed noisy', and `online WPE' were computed using the top center microphone of HL2.
The online WPE (Sect.~\ref{sec:proposed_frontend_dereverb})
was also used with the delay-and-sum (DS) beamforming or the minimum power distortionless response (MPDR) beamforming.
For `online WPE + DS' and `online WPE + MPDR',
we selected one vector from the set of pre-recorded steering vectors given the target directions.
For `online WPE + MPDR', we used an EMA of mixture covariance matrices computed similar to Eq.~\eqref{eq:gamma}.
Thus, its performance affected by the block size and shift size.
The shown WERs for
16 ms $\smash{\left(T^\text{BF} \!=\! 1\right)}$,
64 ms $\smash{\left(T^\text{BF} \!=\! 4\right)}$,
and 256 ms $\smash{\left(T^\text{BF} \!=\! 16\right)}$
were achieved using the optimal $\alpha^\text{BF}$, i.e.,
0.020, 0.100, 0.200, respectively.
The performances of `WPE + MVDR with DNN-based mask estimation' were the best ones shown in \cite{sekiguchi22adaptive}.

\begin{table}[t]
    \renewcommand{\arraystretch}{1.025}
    \setlength\tabcolsep{6pt}         %
    \setlength\aboverulesep{1pt}      %
    \setlength\belowrulesep{1pt}       %
    \vspace{-.5\baselineskip}
    \centering
    \caption{
        Computation times and total latencies [ms] of the \textit{proposed} front end for different $T^\text{BF}$ [frames] (with no overlap).
        Lower computation time is better.
    }
    \label{tab:comp_time_proposed}
    \vspace{.1\baselineskip}
    \begin{tabular}{@{}l*{6}{r}@{}}
        \toprule
        Block size $T^\text{BF}$ [frames] & $1$ & $2$ & $4$ & $8$ & $16$ & $32$ \\
        \midrule
        Block shift [ms] & $16$ & $32$ & $64$ & $128$ & $256$ & $512$ \\
        Computation time [ms] & $6$ & $10$ & $17$ & $30$ & $57$ & $111$ \\
        Total latency [ms] & $22$ & $42$ & $81$ & $158$ & $313$ & $623$ \\
        \bottomrule
    \end{tabular}
\end{table}

\begin{table}[t]
    \renewcommand{\arraystretch}{1.1}
    \setlength\tabcolsep{2.5pt}         %
    \setlength\aboverulesep{1pt}      %
    \setlength\belowrulesep{1pt}       %
    \newcommand{\cHead}[1] { \rotatebox[origin=c]{45}{$\,#1\,$} }
    \newcommand{\cTab}[1] { \cellcolor{black!\xinttheiexpr (100-#1)\relax} \xintifboolexpr {#1<=50} {\textcolor{white}{$#1$}} {$#1$} }
    \newcommand{\cTabBfIt}[1] { \cellcolor{black!\xinttheiexpr (100-#1)\relax} \xintifboolexpr {#1<=50} {\textcolor{white}{$\bm{\mathit{#1}}$}} {$\bm{\mathit{#1}}$} }
    \newcommand{\cTabBf}[1] { \cellcolor{black!\xinttheiexpr (100-#1)\relax} \xintifboolexpr {#1<=50} {\textcolor{white}{$\mathbf{#1}$}} {$\mathbf{#1}$} }
    \newcommand{\cTabIt}[1] { \cellcolor{black!\xinttheiexpr (100-#1)\relax} \xintifboolexpr {#1<=50} {\textcolor{white}{$\mathit{#1}$}} {$\mathit{#1}$} }
    \vspace{-.5\baselineskip}
    \centering
    \caption{
        Average WERs [\%] of the \textit{proposed} system for different $T^\text{BF}$ [frames] and $\alpha^\text{BF}$.
        Lower WER score is better. For visualization purpose, the shading of each cell reflects the WER score.
        The best performance for each $T^\text{BF}$ is in bold type.
        The top performances that are not statistically different from
        the overall best performance indicated by $^\bigstar$ are marked with $^{\openbigstar[.5]}$. \looseness=-1
    }
    \label{tab:wer_proposed}
    \vspace{.1\baselineskip}
    \begin{tabular}{@{}r*{7}{c}@{}}
        & \multicolumn{7}{c}{$\alpha^\text{BF}$} \\
        \cmidrule(lr){2-8}
        \multirow{-2}{*}[6pt]{\rotatebox[origin=c]{90}{$\leftarrow T^\text{BF}$}} & \cHead{0.500} & \cHead{0.200} & \cHead{0.100} & \cHead{0.050} & \cHead{0.020} & \cHead{0.010} & \cHead{0.005} \\
        $32$ & \cTab{15.8}{\color{white} ${\!^{\openbigstarwhite[.5]}}\!$} & \cTabBf{15.2}{\color{white} ${\!^{\openbigstarwhite[.5]}}\!$} & \cTab{16.8} & \cTab{21.8} & \cTab{27.9} & \cTab{38.1} & \cTab{50.6} \\
        $16$ & \cTab{20.1} & \cTabBf{15.0}{\color{white} ${\!^{\openbigstarwhite[.5]}}\!$} & \cTab{15.7}{\color{white} ${\!^{\openbigstarwhite[.5]}}\!$} & \cTab{17.8} & \cTab{23.7} & \cTab{28.1} & \cTab{36.8} \\
         $8$ & \cTab{39.0} & \cTab{18.3} & \cTabBf{14.9}{\color{white} ${\!^{\openbigstarwhite[.5]}}\!$} & \cTab{15.7}{\color{white} ${\!^{\openbigstarwhite[.5]}}\!$} & \cTab{18.7} & \cTab{23.1} & \cTab{27.4} \\
         $4$ & \cTab{93.8} & \cTab{30.9} & \cTab{18.7} & \cTabBf{14.9}{\color{white} ${\!^{\openbigstarwhite[.5]}}\!$} & \cTab{15.8}{\color{white} ${\!^{\openbigstarwhite[.5]}}\!$} & \cTab{19.7} & \cTab{23.0} \\
         $2$ & \cTab{95.4} & \cTab{75.7} & \cTab{30.9} & \cTab{18.8} & \cTabBf{14.8}{\color{white} ${\!^\bigstar}\!$} & \cTab{16.1} & \cTab{20.1} \\
         $1$ & \cTab{98.3} & \cTab{97.2} & \cTab{68.3} & \cTab{30.7} & \cTab{16.8} & \cTabBf{15.4}{\color{white} ${\!^{\openbigstarwhite[.5]}}\!$} & \cTab{16.6} \\
    \end{tabular}
    \vspace{.25\baselineskip}
\end{table}

Tables \ref{tab:comp_time_proposed} and \ref{tab:wer_proposed} show
the computational times and the average word error rates (WERs), respectively, of the proposed system
for different $T^\text{BF}$ (with no overlap) and different $\alpha^\text{BF}$.
The WERs marked with $^{\openbigstar[.5]}$
are not statistically different
(the null hypothesis is accepted at the 95\% confidence level)
from the best WER marked with $^\bigstar$,
i.e., 14.8\% for $\smash{\left(T^\text{BF} \!=\! 2, \alpha^\text{BF} \!=\! 0.020\right)}$.   %
It is worth noting that for $\smash{\left(T^\text{BF} \!=\! 1, \alpha^\text{BF} \!=\! 0.010\right)}$, the WER (i.e., 15.4\%)   %
is statistically the same as the best performance.
It demonstrates that our proposed system can also perform very well even with the low-latency frame-online processing.

The optimal $\smash{\alpha^\text{BF}}$ for each $\smash{T^\text{BF}}$ %
suggests that when a small block size is used,
the front end should rely more on the accumulated statistics and put less importance on the newly acquired data
(cf. Eqs. \eqref{eq:gamma} and \eqref{eq:upsilon}).
Inappropriate $\smash{\alpha^\text{BF}}$ may have detrimental effects,
e.g., $\smash{\alpha^\text{BF}} \!=\! 0.500$ for $\smash{T^\text{BF}} \!\in\! \{1, 2, 4\}$.
Using optimal $\smash{\alpha^\text{BF}}$,
our proposed system outperformed the baseline performances,
including that of mask-based MVDR with DNN \cite{sekiguchi22adaptive}.
It suggests that our estimation of source covariance matrices for deriving the MVDR beamformer was more adaptive in handling the sound scene changes due to the interfering speaker movements.
Accumulating the statistics using EMA seems crucial and may also benefit MVDR with DNN-based mask estimation.
We leave this for future work.

\vspace{-.25\baselineskip}
\section{Conclusion}
\label{sec:conclusion}
\vspace{-.125\baselineskip}

This paper proposes a practical approach to the enhancement of adaptive speech with low latency and high performance.
The system operates an online MVDR beamforming on the front end that
adopts the posterior distribution obtained by the back-end BSS based on FastMNMF.
Future works include 
considering scenarios with continuously moving sources and
automating hyperparameter tuning for $\smash{\alpha^\text{WPE}}$, $\smash{\alpha^\text{BF}}$, and $\smash{\alpha^\text{BSS}}$.

\clearpage
\begingroup
\newcommand{\myfontsize}{\fontsize{9.7}{10}\selectfont}
\def\baselinestretch{.99}\let\normalsize\myfontsize\normalsize
\bibliographystyle{./IEEEbib_abrv}
\bibliography{./IEEEabrv,./MYabrv,./refs}
\endgroup

\end{document}